\documentstyle[preprint,aps]{revtex}
\begin{document}
\title{Stable autosolitons in dispersive media with saturable gain and absorption}
\author{Boris A. Malomed$^{a}$, Andrei G. Vladimirov$^{b}$, Galina V. Khodova$^{c}$,
and Nikolay N. Rosanov$^{c}$}
\address{$^{a}$Department of Interdisciplinary Studies, Faculty of Engineering, Tel\\
Aviv University, Tel Aviv 69978, Israel\\
$^{b}$St.Petersburg State University, Physics Faculty, 1 Ulianovskaya st,\\
198904 St.Petersburg, Russia\\
$^{c}$Institute for Laser Physics, Birzhevaya line 12, 199034 St.\\
Petersburg, Russia}
\maketitle

%\date{\today}
%\pacs{}

\newpage

%\begin{abstract}

\begin{center}
ABSTRACT
\end{center}

We introduce the simplest one-dimensional model of a dispersive optical
medium with saturable dissipative nonlinearity and filtering (dispersive
loss) which gives rise to stable solitary pulses (autosolitons). In the
particular case when the dispersive loss is absent, the same model may also
be interpreted as describing a stationary field in a planar optical
waveguide with uniformly distributed saturable gain and absorption. In a
certain region of the model's parameter space, two coexisting solitary-pulse
solutions are found numerically, one of which may be stable. Solving the
corresponding linearized eigenvalue problem, we identify stability borders
for the solitary pulses in their parametric plane. Beyond one of the
borders, the symmetric pulse is destroyed by asymmetric perturbations, and
at the other border it undergoes a Hopf bifurcation, which may turn it into
a breather.

%\end{abstract}

\newpage

\section{Introduction}

``Autosolitons'' are robust localized pulses in models combining
conservative and dissipative nonlinear and dispersive terms \cite{book}.
They occur and find important applications in nonequilibrium plasmas \cite
{Stenflo,book} and semiconductors \cite{book}, in hydrodynamics (Poiseille
flow \cite{Poiseille} and traveling-wave convection \cite{convection}), and
in nonlinear optics \cite{Rosanov}-\cite{Petersburg2}.

The simplest model which gives rise to solitary pulses that may be
interpreted as autosolitons is the cubic complex Ginzburg-Landau (GL)
equation, 
\begin{equation}
E_{z}=E+(1-ic_{1})E_{\tau \tau }-(1+ic_{2})|E|^{2}E,  \label{cubic}
\end{equation}
which is written in the standard ``optical'' notation \cite{optics}, i.e., $%
E $ is the local amplitude of the electromagnetic wave, $z$ (to be treated
as the evolutional variable) is the propagation distance, and $\tau =t-z/V_{%
{\rm gr}}$ is the ``local time'', $t$ and $V_{{\rm gr}}$ being the physical
time and the mean group velocity of the carrier wave. Further, $c_{1}$ and $%
c_{2}$ are coefficients of the (chromatic) dispersion and (Kerr)
nonlinearity, while the terms the coefficients in front of which are
normalized to be $1$ account for, respectively, linear gain, dispersive
losses (spectral filtering), and nonlinear losses (two-photon absorption).
Equation (\ref{cubic}) always has a single exact solitary-pulse solution 
\cite{Poiseille,Stenflo}; however, this solution is always unstable, as its
background, the trivial solution $E=0$, is obviously unstable because of the
presence of the linear gain.

The simplest possibility to modify the model so that to let it generate {\em %
stable} solitary pulses is to convert it into the quintic GL equation \cite
{quintic}, 
\begin{equation}
E_{z}=-E+(1-ic_{1})E_{\tau \tau }+(1+ic_{2})|E|^{2}E-(\Gamma
+ic_{3})|E|^{5}E,  \label{quintic}
\end{equation}
where the linear gain and cubic loss are replaced, respectively, by linear
loss and cubic gain, $\Gamma >0$ is the coefficient of quintic loss, and $%
c_{3}$ accounts for a possible quintic correction to the Kerr effect. In
this case, the trivial solution is stable, and two solitary-pulse solutions,
one unstable and one stable, may coexist at fixed values of all the
parameters in eq. (\ref{quintic}).

However, the quintic GL equation is a phenomenological model, and it would
be very desirable to find more realistic models allowing for the existence
of stable solitary pulses. One possibility is to consider a model of a {\em %
dual-core} nonlinear optical fiber, in which the linear gain, dispersion,
filtering, and Kerr (purely cubic) nonlinearity are present in one (active)
core, while the other (passive) one, linearly coupled to the active core,
has only linear loss \cite{Atai1}. Thus, the model consists of a cubic GL
equation linearly coupled to the second, purely linear, equation. It is easy
to select parameters of the model so that to provide for the stability of
the trivial solution. Then, two {\em exact} solitary-pulse solutions
(autosolitons) can be found, following the pattern of the exact solution to
the cubic GL equation, and direct numerical simulations clearly show that
one of the two exact solitary pulses may be stable \cite{Atai1}. The
simulations have revealed nontrivial autosoliton's stability borders in the
parameter space of the model (``nontrivial'' implies that the borders are
different from obvious stability conditions for the zero background).
Moreover, interactions between the stable autosolitons have also been
simulated in detail in this model \cite{Atai2}.

\section{The model}

In this work, we aim to demonstrate that stable optical autosolitons can
also be easily found in a model similar to that introduced in \cite{RF92},
which combines {\em saturable} gain and {\em saturable} absorption in a
lasing medium, while the Kerr nonlinearity may be completely neglected.
Thus, the nonlinearity in the model is purely dissipative. The gain and loss
coefficients in the model are to be chosen so that its linearized version
has no gain, in order to provide for the stability of the zero background.
The presence of the linear loss suggests that linear dispersive loss, i.e.,
the filtering (diffusion-like) term should also be included.

All the ingredients of the saturable model discussed above are dissipative
(considering gain as negative dissipation). However, from the experience
accumulated in the studies of the quintic GL equation in the
strong-dissipation limit \cite{quintic2}, it is known that, while
solitary-pulse solutions may exist in a purely dissipative model, they can
never be stable \cite{ZBL}. The presence of one, at least, conservative term
(which may enter with a small coefficient \cite{quintic2}) is necessary to
provide for the stability of the corresponding autosoliton. In optical
media, this role may be very naturally played by the dispersion.

Thus, the model combining saturable gain, saturable absorption, dispersion
and (not necessarily) filtering takes the form (cf. similar models for the
lasers with saturable absorbers \cite{RF92,Petersburg1,Petersburg2,FVKR00}): 
\begin{equation}
E_{z}-ie^{-i\theta }E_{\tau \tau }=\left( -1+\frac{g_{0}}{1+I/I_{g}}-\frac{%
a_{0}}{1+I/I_{a}}\right) E,\,\,I\equiv |E|^{2},  \label{model}
\end{equation}
where $g_{0}$ and $a_{0}$ are positive gain and absorption coefficients in
the linear approximation, and $I_{g}$ and $I_{a}$ are saturation intensities
for the gain and absorption (the linear loss coefficient is normalized to be 
$1$, cf. eq. (\ref{quintic})). The real coefficient $\theta $ characterizes
a relation between the dispersion and filtering (diffusion) coefficients,
which are proportional, respectively, to $\cos \theta $ and $\sin \theta $.
By means of an obvious rescaling, we may set $I_{a}=1$, then all the
remaining parameters are fully independent, and their number cannot be
further reduced.

The underlying condition of the stability of the zero background ($E=0)$
takes the form 
\begin{equation}
g_{0}<1+a_{0}\,,  \label{zero}
\end{equation}
which must be supplemented by $g_{0}>1$, as otherwise the model can never
provide for an effective gain. The positiveness of the effective filtering
coefficient imposes another necessary condition, $0\leq \theta \leq \pi $.
In fact, the range of the parameter $\theta $ can be restricted to 
\begin{equation}
0\leq \theta <\pi /2\,,  \label{theta}
\end{equation}
as the region $\pi /2<\theta \leq \pi $ can be mapped into $0\leq \theta
<\pi /2$ by means of complex conjugation, $E\rightarrow E^{\,\ast }$. The
value $\theta =\pi /2$ is excluded, as it corresponds to the purely
dissipative model that cannot give rise to stable pulses.

In the particular case $\theta =0$ (no filtering), the same model with the
temporal variable $\tau $ replaced by the transverse coordinate $x$ has an
alternative interpretation in nonlinear optics: it may describe a stationary
field in a planar waveguide \cite{optics} with the saturable gain and
absorption uniformly distributed in it. In that case, the term $iE_{xx}$
accounts for the diffraction in the waveguide.

\section{Autosolitons and their stability in the saturable model}

An autosoliton (solitary-pulse) solution to eq. (\ref{model}) is sought for
in an obvious form, $E(z,\tau )=\exp (i\alpha z)\cdot {\cal E}_{0}(\tau )$,
where $\alpha $ is the propagation constant, and ${\cal E}_{0}(\tau )$ is a
complex {\em even} function vanishing at $|\tau |\rightarrow \infty $ and
determined by an equation 
\begin{equation}
\frac{d_{0}^{2}{\cal E}}{d\tau ^{2}}=ie^{i\theta }\left( -1+i\alpha +\frac{%
g_{0}}{1+|{\cal E}_{0}|^{2}/I_{g}}-\frac{a_{0}}{1+|{\cal E}_{0}|^{2}/I_{a}}%
\right) {\cal E}_{0}\,.  \label{U}
\end{equation}
As is well known \cite{quintic2,Atai1}, an autosoliton will have no chance
to be stable if only {\em one} solitary-pulse solution exists at given
values of the parameters. Indeed, due to the condition (\ref{zero}), the
system has a trivial attractor, $E=0$. If there is also a nontrivial
attractor in the form of an autosoliton, there must simultaneously exist an
unstable solitary-pulse solution that plays the role of a {\it separatrix}
between the attraction basins of the two attractors. It is well known too
that the two solitary-pulse solutions may undergo a bifurcation at some
critical point, where they merge and disappear, so that no autosoliton
exists past this point.

In exact accordance with these expectations, numerical integration of eq. (%
\ref{U}) reveals that solitary pulses exist only in a pair in some
parametric region, and do not exist at all in other regions. As a typical
illustration, in Fig. 1 we display the autosoliton's propagation constant as
a function of $\theta $ at fixed values of the other parameters. In the
particular case shown in this figure, the autosolitons actually exist at $%
0\leq \theta <0.34$, i.e., inside a relatively small part of the formally
available region (\ref{theta}), which implies that the saturable gain cannot
compensate filtering losses when they are too strong (see also below). Note
that, despite the large ratio $I_{g}/I_{a}=10$ in the case shown in Fig. 1,
we cannot use an approximation with $I_{g}=\infty $: as it is follows from a
simple consideration of eq.(\ref{model}), in this limit the model either can
provide no effective gain at all, if $g_{0}<1$, or it will blow up (be
unstable at $|E|^{2}\rightarrow \infty $) in the opposite case. Above a
certain threshold value of the linear gain parameter $g_{0}$ the upper and
lower autosoliton branches do not merge any longer with the increase of $%
\theta $ and, hence, the saddle-node bifurcation shown in Fig. 1 disappears
(see Fig. 2). In Fig. 2, which corresponds to $g_{0}=2.11$, T is the limit
point for the stable autosoliton branch. When approaching this point from
the left the width of the autosoliton solution tends to infinity.

For the study of the stability of the stationary autosolitons, we linearized
the full equation (\ref{model}) near the stationary solution, assuming $%
E(z,\tau )=\exp (i\alpha z)\cdot \left[ {\cal E}_{0}(\tau )+\exp \left(
\gamma z\right) {\cal E}_{1}(\tau )\right] $, where ${\cal E}_{1}(\tau )$ is
an eigenmode of the infinitesimal perturbation, and $\gamma $ is the
corresponding instability growth rate. Because the unperturbed solution $%
{\cal E}_{0}(\tau )$ is even, the resultant linear eigenvalue problem for $%
\gamma $ can be solved separately for even (symmetric) and odd
(antisymmetric) eigenmodes ${\cal E}_{1}$ \cite{VFKKR99}.

Numerical results for the corresponding eigenvalues are presented in Fig. 3
(for the same values of the parameters as in Fig. 1). As is it obvious from
the figure, all the upper branch of the stationary solutions from Fig. 1 is
unstable against symmetric perturbations, while all the lower branch is
stable against them. However, a part of the stationary solutions belonging
to the lower branch is destabilized by antisymmetric perturbations (this
part of the lower branch is dotted in Fig. 1). It is noteworthy that the
antisymmetric eigenmodes related to the upper branch in Fig. 1 undergo a
bifurcation at $\theta \approx 0.22$, therefore the corresponding curve $%
\gamma _{{\rm as}}(\theta )$ in Fig. 3b has several parts, marked by the
numbers $2$, $3$, and $4$ (mark $1$ is reserved for the smooth curve $\gamma
_{{\rm as}}(\theta )$ corresponding to the lower-branch autosolitons).

To present the stability results in a possibly most general and compact
form, we continued the numerical analysis, varying $\theta $ and,
additionally, the gain parameter $g_{0}$, while for the absorption
coefficient and the ratio of the saturation intensities the same fixed
values were kept as in Figs. 1,2, and 3, i.e., $a_{0}=2$ and $I_{g}/I_{0}=10$%
. This way to vary the parameters has a clear physical sense, as in the
experiment the loss factor and saturation intensity ratio are both fixed for
a given setup, while the gain can be readily adjusted changing the pump
power.

The region in the parametric plane ($g_{0}$,$\theta $) where the autosoliton
is stable\ according to the numerical solution of the linearized eigenvalue
problem is shown in Fig. 4. Above the upper border AS, the autosoliton loses
its stability to antisymmetric perturbations, similarly to what was
discussed in detail above for the particular case $g_{0}=2.06$. Another
upper border, T, corresponds to a bifurcation set where the stable
autosoliton branch terminates as it is shown in Fig. 2. The lower border, H,
starting at $g_{0}\approx 2.094$, is a new one: at this border, the
autosoliton loses stability against a perturbation eigenmode with a complex
instability growth rate, corresponding to a {\it Hopf bifurcation} that is
going to transform the stationary autosoliton into a breather (vibrating
autosoliton). A detailed study of the breather is beyond the scope of the
present work.

Note that the autosolitons do not exist at all at $g_{0}<2.026$, which can
be easily explained by the fact that, when the gain is too weak, it cannot
provide for the balance with loss, necessary for the existence of a
stationary pulse. It is also noteworthy that the upper stability border is
going up very steeply with the increase of $g_{0}$, which may be
qualitatively realized too: an excessive gain makes it possible to
compensate extra filtering losses proportional to $\sin \theta $. Besides
that, it was argued above that the region $\pi /2<\theta \leq \pi $ is fully
symmetric to the region (\ref{theta}); from here it follows that, in order
to comply with the symmetry, the upper border must turn back at $\theta =\pi
/2$, which explains its steep ascent.

\section{Conclusion}

We have described the simplest model of a dispersive optical medium (which,
in the general case, includes dispersive loss too) with saturable
dissipative nonlinearity, which gives rise to stable solitary pulses. If the
dispersive loss is absent, the same model may also be interpreted as
describing a stationary field in a planar optical waveguide with uniformly
distributed saturable gain and absorption. In a certain parametric region,
two coexisting solitary-pulse solutions were found numerically, one of which
may be stable. Numerical solution of the corresponding linearized eigenvalue
problem has determined actual stability borders for the autosoliton. Beyond
one of the borders, the symmetric autosoliton is destroyed by asymmetric
perturbations, while at the other border it undergoes a Hopf bifurcation,
which may turn it into a breather.

\newpage

\section*{FIGURE CAPTIONS}

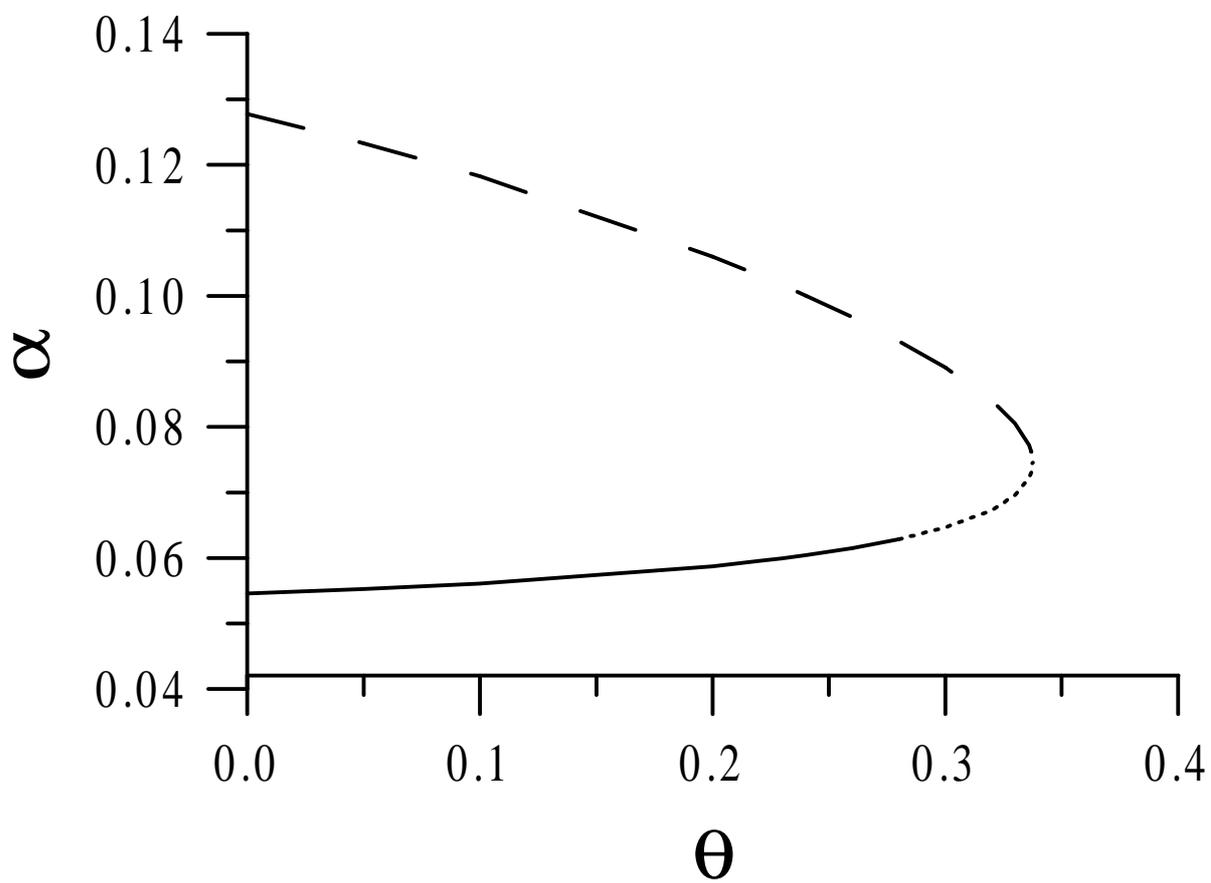
Fig. 1. The autosoliton's propagation constant vs. the parameter $\theta $
controling the ratio between the dispersion and filtering (diffusion)
coefficients, that may vary in the interval (\ref{theta}). In this figure
and in Fig. 2 below, the other parameters are $g_{0}=2.06$, $a_{0}=2$, and $%
I_{g}/I_{a}=10$. Two branches of the autosoliton solutions correspond to the
upper and lower curves in this plot. The whole upper branch, and the dotted
part of the lower one correspond to unstable autosolitons, see below.

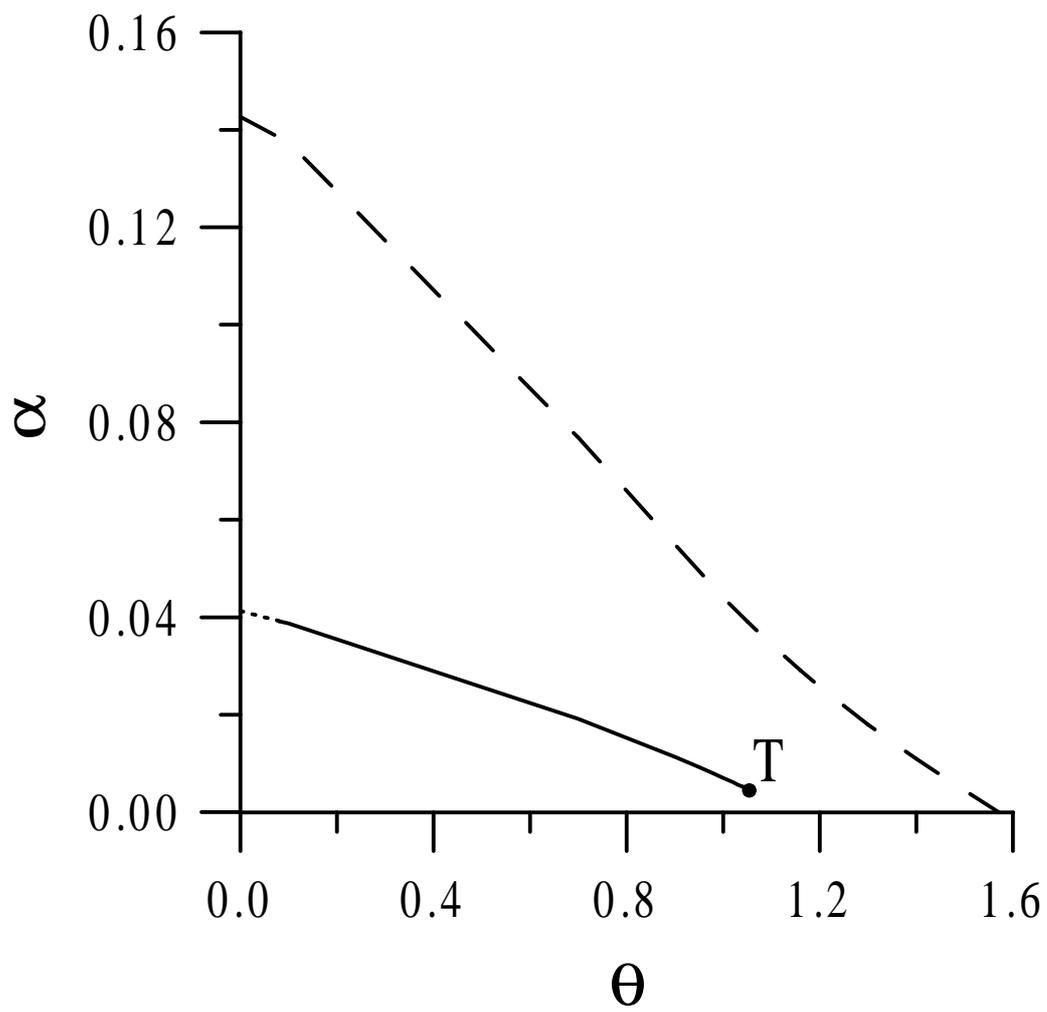
Fig. 2. Same as in Fig. 1, but for $g_{0}=2.11$. The upper autosoliton
branch is always unstable and terminates at the point $\alpha =0$, $\theta
=\pi /2$, which corresponds to the unstable autosoliton of the purely
dissipative model. The lower branch becomes unstable via a Hopf bifurcation
for small values of the parameter $\theta $ (dotted line). This branch
terminates at the point T.

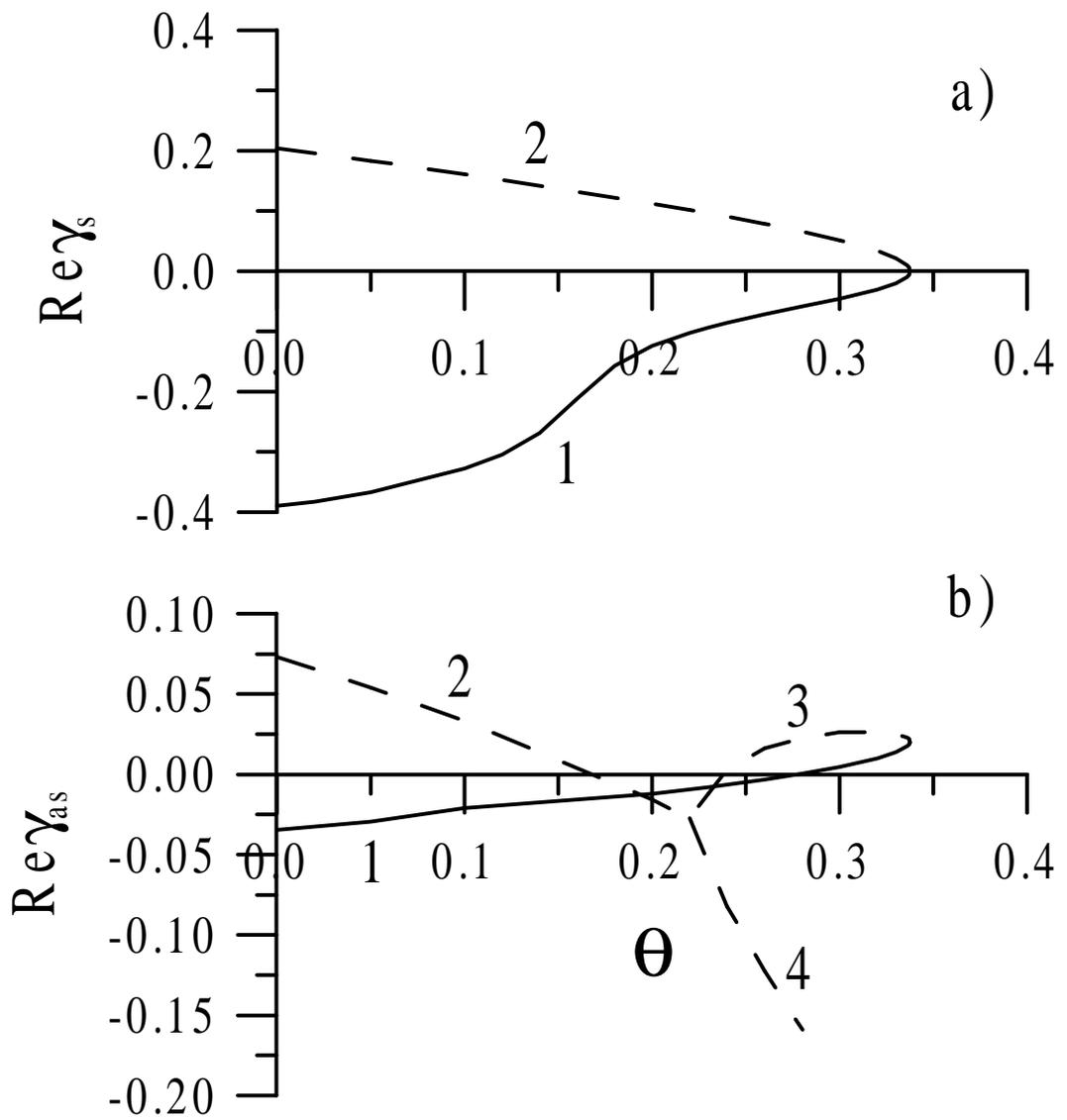
Fig. 3. The instability growth rates $\gamma _{{\rm s}}$ and $\gamma _{{\rm %
as}}$ for the symmetric (a) and antisymmetric (b) infinitesimal
perturbations of the stationary autosolitons vs. $\theta $. The dashed and
continuous curves $\gamma (\theta )$ pertain, respectively, to the upper and
lower branches in Fig. 1.

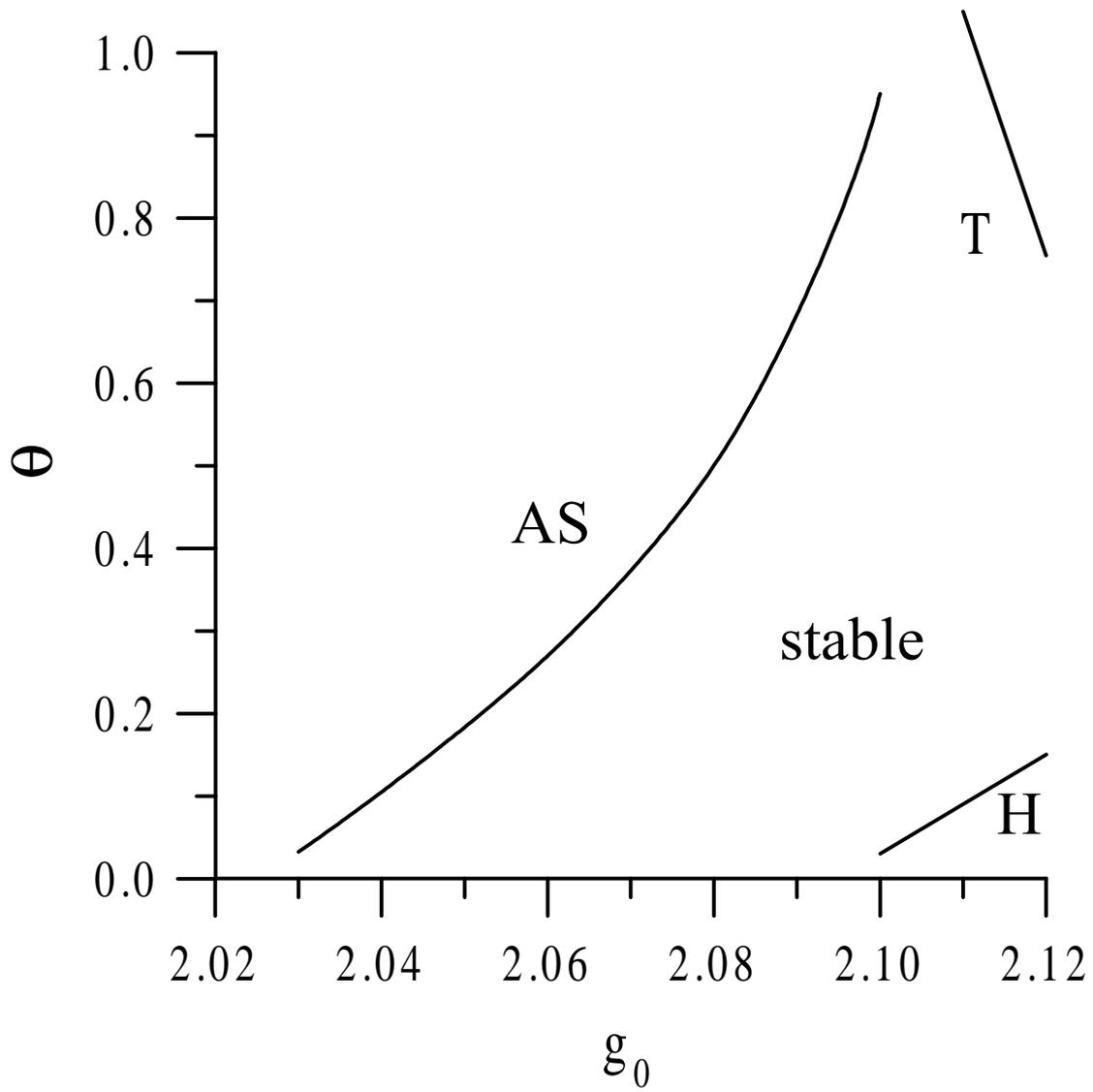
Fig. 4. Stability borders for the autosoliton in the ($g_{0}$,$\theta $)
plane. The other parameters are $a_{0}=2$ and $I_{g}/I_{a}=10$.

\end{document}